\documentstyle[12pt,epsfig,epsf]{article}
\def\beq{\begin{equation}}
\def\eeq{\end{equation}}
\def\bea{\begin{eqnarray}}
\def\eea{\end{eqnarray}}
\def\bq{\begin{quote}}
\def\eq{\end{quote}}

\def\gappeq{\mathrel{\rlap {\raise.5ex\hbox{$>$}}
{\lower.5ex\hbox{$\sim$}}}}

\def\lappeq{\mathrel{\rlap{\raise.5ex\hbox{$<$}}
{\lower.5ex\hbox{$\sim$}}}}

\def\Toprel#1\over#2{\mathrel{\mathop{#2}\limits^{#1}}}

\newcommand{\nd}[1]{/\hspace{-0.5em} #1}

 \def\gappeq{\mathrel{\rlap {\raise.5ex\hbox{$>$}}
{\lower.5ex\hbox{$\sim$}}}}
 
\def\lappeq{\mathrel{\rlap{\raise.5ex\hbox{$<$}}
{\lower.5ex\hbox{$\sim$}}}}
    
\begin{document}
\pagestyle{empty}
\begin{flushright}
CERN-TH/2001-134\\
ACT-05-01 \\
CTP-TAMU-16/01 \\
hep-th/0105206\\
\end{flushright}
\vspace*{5mm}
\begin{center}
{\bf \$TRING THEORY AND AN ACCELERATING UNIVERSE} \\
\vspace*{1cm} 
{\bf John ELLIS$^{a)}$, N.E. MAVROMATOS$^{b)}$
and D.V. NANOPOULOS$^{c)}$}

a) Theoretical Physics Division, CERN, CH 1211 Geneva 23 \\

b) Department of Physics, Theoretical Physics, King's College London, 
Strand, London WC2R 2LS, U.K. \\ 

c) Department of Physics, Texas A \& M University, 
College Station, TX~77843, USA; \\
Astroparticle Physics Group, Houston
Advanced Research Center (HARC), 
Mitchell Campus,
Woodlands, TX~77381, USA; \\
Chair of Theoretical Physics, 
Academy of Athens, 
Division of Natural Sciences, 
28~Panepistimiou Avenue, 
Athens 10679, Greece

\vspace*{2cm}  
{\bf ABSTRACT} \\ \end{center}
\vspace*{5mm}

An accelerating Universe can be accommodated naturally within non-critical
string theory, in which scattering is described by a superscattering
matrix \$ that does not factorize as a product of $S$- and
$S^\dagger$-matrix elements and time evolution is described by a modified
Liouville equation characteristic of open quantum-mechanical systems. We
describe briefly alternative representations in terms of the stochastic
Ito and Fokker-Planck equations. The link between the vacuum energy and
the departure from criticality is stressed. We give an explicit example in
which non-marginal \$tring couplings cause a departure from criticality,
and the corresponding cosmological vacuum energy relaxes to zero {\it \`a
la} quintessence. 

\vspace*{3cm}

\begin{flushleft} CERN-TH/2001-134 \\
May 2001
\end{flushleft}
\vfill\eject

\setcounter{page}{1}
\pagestyle{plain}

Conceived parthenogenically to model the $S$ matrix of the strong
interactions~\cite{GV}, string theory aspires to apotheosis as a quantum
theory of gravity~\cite{SS}. To demonstrate its worthiness for this
exalted status, string theory must accomplish successfully at least four
Labours. The first of these to be accomplished was the cleansing of the
Augean stables of perturbative quantum-gravitational divergences in the
weak-field limit~\cite{GS}.  The second was to slay the Gorgon of quantum
black holes by identifying the quantum states that quantify their
entropy~\cite{entropy}. However, looking at the Gorgon's face is still
problematic: can one measure all these states in a realistic experiment?
If not, the black-hole information-loss problem is reformulated rather
than exorcised~\cite{EMNreview}. 

The third Labour is to tame the microscopic quantum fluctuations in the
foamy space-time background. We have argued that the most appropriate
framework for describing particles in interaction with an unmeasured (and
unmeasurable) recoiling background is an effective non-critical string
theory~\cite{EMNreview}, which allows for information loss and entropy
growth.  Scattering might be describable asymptotically by a
superscattering matrix \$, that is not factorizable as a simple product of
the $S$ matrix and its hermitean conjugate $S^\dagger$~\cite{Hawking}.
{\it In order to achieve its immortal apotheosis, string theory may need
to shed its mortal $S$-matrix origins and criticality to become
non-critical \$tring theory.}
In many physical applications, a sub-asymptotic time-evolution equation is
useful, and we advocate~\cite{EMNdollar} a generalized quantum Liouville
equation for the density matrix~\cite{ehns}:
\beq
\dot\rho = i \bigg[ H,\rho\bigg] + \nd{\delta H} \rho
\label{one}
\eeq
where $\nd{\delta H}$ represents the interaction of the observed system with a
quantum-gravitational background `environment' associated with
microscopic event horizons. 

The fourth Labour is to describe quantum cosmology. For this, the first
requirement was to formulate string theory in a suitable
Robertson-Walker-Friedman background.  An early approach to this problem
showed how a classical non-critical string theory could be used to
accommodate a time-dependent scale factor~\cite{aben}. Subsequently, the
issue of entropy generation during a primordial inflationary epoch was
addressed, and a framework based on quantum non-critical \$tring theory
was
proposed~\cite{emninfl}. More recently, this approach was extended to
asymptotically large times, and this \$tring theory was used to suggest
that relaxation of the quantum-gravitational metric might make a
time-dependent contribution to the cosmological vacuum
energy~\cite{cosmoconst}, sufficient to accelerate the Universe even in
the absence of any other contributions to the vacuum energy. This
suggestion was linked intimately to the entropy growth intrinsic to
\$tring theory, that is generically associated with event horizons - in
this case macroscopic.

The observational evidence for an accelerating Universe is becoming
overwhelming, with cosmic microwave background measurements~\cite{cmb}
converging on a flat Universe containing a low baryon density consistent
with predictions based on cosmological nucleosysnthesis -- conferring
credibility on the parameter fits -- and vacuum energy with a density
about the two-thirds of the critical flat-space value -- consistent with
interferences from large-scale structure and high-redshift
supernovae~\cite{perlmutter}.

This is a cause for concern within critical-string theory~\cite{banks}: `a
challenge for
string theory as presently defined'~\cite{susskind} and `the current sets
of concepts in string theory will not be sufficient to give a {\it
coherent} description of our Universe' (our
italics)~\cite{susskind}, or `quintessence, very much like a
cosmological constant, presents a serious challenge for string
theory'~\cite{fischler}.

The purpose of this paper is to recall~\cite{EMNreview} that non-critical
\$tring theory provides a suitable {\it decohering} framework for a
quantum description of the Universe, outlining its observables, and
discussing the dynamical equations they obey. We extend our previous
discussion of vacuum energy within this framework. We develop an explicit
example in which non-zero $\beta$ functions for \$tring $\sigma$-model
couplings cause a departure from criticality and cosmological vacuum
energy that relaxes to zero {\it \`a la} quintessence. 

Critical string theory is described by a conformal two-dimensional field theory
($\sigma$ model) on the world sheet. Non-critical \$tring
theory~\cite{ddk} is
formulated in terms of non-conformal renormalizable world-sheet field theories,
described by an action 
\beq
A[g] = A[g^*] + \int d^2z \sqrt\gamma ~~ g^i V_i
\label{two}
\eeq
where the $\{g^i\}$ denote the couplings of massless background field
deformations $\{V_i\}$ that are not exactly marginal, $\{g^*\}$ represents
the
equilibrium  conformal background around which we perturb, and $\gamma\equiv
e^\phi \gamma^*$, where $\phi$ is the world-sheet Liouville field 
and $\gamma^*$
denotes a reference world-sheet metric. We interpret $\phi$ as a world-sheet
renormalization scale~\cite{EMNreview}, so that non-conformal scale
dependence is compensated by
Liouville dressing: the gravitationally-dressed operators $V_i \rightarrow
[V_i]_\phi$ are exactly marginal in a renormalization-group sense. We adopt the
Wilson renormalization-group picture, according to which non-trivial scaling in
the effective theory consisting only of the massless modes $\{g^i\}$ 
follows from
integrating out the other degrees of freedom. {\it In an accelerating Universe,
these would include the degrees of freedom that disappear through the event
horizon}. 

We have argued~\cite{EMNreview} that they also include the quantum degrees of
freedom associated with microscopic topological fluctuations in the
space-time foam. In the case of a two-dimensional space-time black hole
model, it was shown explicitly that the vertex operator for a massless
`tachyon' field alone is {\it not} exactly marginal, but only becomes so
in combination with a vertex operator for a higher-level non-local degree
of freedom associated with the black hole~\cite{whair}
We further argued
that the quantum state of a two-dimensional string black hole is
characterized by the quantum numbers associated with a complete set of
such non-local vertex operators, that we termed $W$ hair~\cite{whair}. We
conjectured
that this type of characterization of a quantum black hole could be
extended to four dimensions, and just such a solitonic understanding of
the quantum states of four-dimensional black holes was subsequently
attained~\cite{entropy}. However, in our view, observable physics is
obtained by integrating out the non-local vertex operators, leaving behind
a non-conformal effective theory for the massless mode, that is {\it not}
exactly marginal~\cite{qmv}. 

What kind of animal is this effective theory? Critical string theory has a
Lorentz-invariant $S$ matrix, but we cannot expect the same for a
non-critical \$tring theory~\cite{EMNreview,adler}. This obeys a
renormalization-group equation that expresses its dependence on the
Liouville field $\phi$. We identify~\cite{EMNreview} 
the (zero mode) of $\phi$ with the
target time variable. This identification is supported by (i) the fact
that \$tring theory is super-critical in general~\cite{EMNreview}, 
giving
$\phi$ a negative metric~\cite{aben}:  $(-ve) \int d^2z~~\partial\phi
\bar\partial\phi$, (ii) the recovery of the conventional Lorentz-invariant
$S$-matrix description on the critical limit, and (iii) the recovery of
the two-dimensional black-hole metric in an explicit soluble non-critical
example~\cite{whair}. Upon the identification of $\phi$ with time, the
renormalization-group equations become the dynamical equations governing
the time evolution of observables.

\begin{figure}[htb]

\begin{center} 
\epsfig{figure=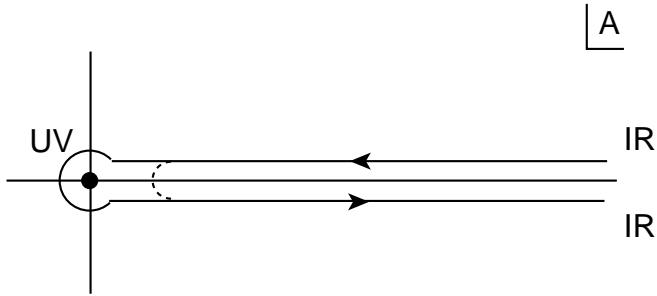}
\hspace{4.0in}
\end{center} 
\caption{{\it Contour
of integration for a proper definition of the path 
integration for the Liouville field, where
the quantity $A$ denotes the (complex) world-sheet area. 
This is known in the literature as the Saalschutz contour,
and has been used in
conventional quantum field theory to relate dimensional
regularization to the Bogoliubov-Parasiuk-Hepp-Zimmermann
renormalization method. Upon the interpetation of the 
Liouville field with target time, this curve
resembles closed-time-paths in non-equilibrium field theories.} }
\label{liouvillecurve}
\end{figure}

We recall briefly our formal arguments~\cite{adler} supporting the point
of view that it is impossible to define a unitary $S$ matrix in Liouville
strings, with the only well-defined object being the \$ matrix.  This
follows from the fact that an $N$-point world-shet correlation function of
vertex operators in Liouville strings, ${\cal F}_N \equiv <V_{i_1} \dots
V_{i_N}>$, where $<\dots>$ signifies a world-sheet expectation value in
the standard Polyakov treatment, transforms under infinitesimal Weyl
shifts of the world-sheet metric $\gamma$ in the following
way~\cite{adler}:
\begin{equation}
\delta_{\rm weyl} {\cal F}_N \, = \, 
\left[ \delta_0
+ {\cal O}\left(\frac{s}{A}\right)\right]{\cal F}_N
\label{weyl}
\end{equation}
where the standard part $\delta_0$ of the variation
involves a sum over 
the conformal dimensions $h_i$ of the operators $V_i$ that is independent
of the world-sheet area $A$ (whose logarithm is the
world-sheet zero mode of the Liouville field). 
The quantity $s$ is the sum of 
the gravitational anomalous dimensions~\cite{ddk}:
\begin{equation}
s=-\sum_{i=1}^{N}\frac{\alpha_i}{\alpha} - \frac{Q}{\alpha}~,
~\alpha_i=-\frac{Q}{2}+ \frac{1}{2}\sqrt{Q^2 + 4(h_i-2)},
~\alpha=-\frac{Q}{2}+ \frac{1}{2}\sqrt{Q^2 + 8}
\label{gad}
\end{equation}
where $Q$ is the central-charge deficit, that is non-zero for
non-critical \$string. 

From (\ref{weyl}), and upon the identification of the logarithm of the
world-sheet area ${\rm ln}A$ with the target time $t$~\cite{EMNreview}, one
observes that such a correlator cannot represent a unitary $S$-matrix
element in target space, as a result of the explicit $A-$ (i.e. time $t-$)
dependence. Nevertheless, according to the analysis
in~\cite{EMNreview,adler},
it is
clear that one can construct a well-defined quantity free from
such world-sheet area ambiguities.  However, this is a \$ matrix that is
not factorizable into a product of $S$ and $S^\dagger$. This results from
the way one performs the world-sheet path integration over the Liouville
mode $\int D\phi (\dots )$ in a first-quantized approach to string theory. 
This integration can be done via a steepest-descent
method~\cite{kogan,EMNreview,adler}, along the curve in
Fig.~\ref{liouvillecurve}, which is reminiscent of the closed time-like
path used in non-equilibrium field theories~\cite{ctp}. Close to the
ultraviolet fixed point on the world sheet, i.e., in the limit when $A \to
0$, there are divergences. These are responsible for the lack of
factorization of the \$ matrix in this case. The definition of the
Liouville path integral over such a curve is responsible, in this
formalism, for the appearance of a density-matrix rather than a
wave-function interpretation of the world-sheet partition function of the
Liouville string, in full agreement with its non-equilibrioum
(open-system) nature.

This impossibility of defining a consistent unitary $S$-matrix element in
Liouville string, only a \$ matrix, provides an argument that
non-critical string theory may be the key~\cite{cosmoconst} to a
resolution of the accelerating Universe puzzle, where the presence of a
cosmological (particle) horizon seems to prevent a consistent definition
of a unitary $S$ matrix~\cite{banks,susskind,fischler}.

We now return to our discussion of  
qualitative features of Liouville (effective) string theory.
The state of the observable system may be characterized by a density matrix
$\rho$. The full density matrix of string theory might in principle be 
a pure-state
density matrix $\rho =|\Psi > < \Psi |$. However, in our
\$tring
picture $|\Psi > = |\psi, \tilde\psi > $, where $|\psi >$ denotes a state
of the
observable system and $|\tilde\psi >$ the unobserved degrees of freedom that are
integrated out in the renormalization-group approach. The reduced observable
density matrix
$\rho\equiv \int d\tilde\psi ~~ \rho = \int d \tilde\psi | \psi, \tilde\psi > <
\tilde\psi, \psi|$ will in general be mixed as a result of entanglement
with the
unobserved degrees of freedom $|\tilde\psi >$, e.g., those that disappear across
the macroscopic event horizon in an accelerating Universe, or across a
microscopic event horizon in a model of space-time foam.

The renormalization-group equation for the reduced density matrix $\rho$ is easy
to derive and can be cast in the form (\ref{one}), with an explicit form for the
non-Hamiltonian operator $\nd{\delta H}$ in terms of the non-conformal
field couplings $g^i$~\cite{EMNreview}:
\beq
\nd{\delta H} = \beta^i G_{ij}g^j
\label{three}
\eeq
where the $\beta^i$ are the non-trivial renormalization functions of the
$g^i$, and
$G_{ij}$ is a suitable metric in the space of these couplings~\cite{EMNreview}.
An equation of the
form (\ref{one}) is familiar from the theory of quantum-mechanical systems in
interaction with an environment, that is provided in our case by the modes
$|\tilde\psi >$ that disappear across the event horizon (\ref{three}).

It is natural to ask whether one can write down a wave equation for the
observable subsystem $|\psi >$, and an answer is provided by the theory of open
quantum-mechanical systems. One may represent (\ref{one}, \ref{two}) as
a stochastic Ito process for $|\psi >$~\cite{gisin}:
\bea
Q|d\psi > &=& 
        -i  H |\psi > + \sum_e (<B^\dagger_e>_\psi B_e - {1\over 2}~
B^\dagger_e
B_e \nonumber \\
&& -{1\over 2} <B^\dagger_e>_\psi <B_e>_\psi) |\psi> dt + (B_e -
<B_e>_\psi) |\psi > d
\xi_e)
\label{four}
\eea
where $H$ is the Hamiltonian of the subsystem $|\psi >$, and the $B_e,
B^\dagger_e$
are `environment' operators that may be defined as appropriate `square
roots'
of the various partitions of the operator $\beta^i G_{ij} g_j$. 
The $d\xi_e$ are
complex differential random variables associated with stochastic Wiener or
Brownian processes, which represent the environmental effects that are averaged
out by a low-energy local observer.

In the cosmological context, we have proposed applying this formalism to the
problem of inflation~\cite{emninfl}. 
The key issue there may be regarded as the generation of a
large amount of entropy. Whenever one has entanglement with an environment over
which one integrates, as in the non-critical \$tring
formalism, one  necessarily encounters entropy  growth. In the
renormalization-group formalism (\ref{one}, \ref{two}), there is a simple
formula for the rate of growth of the entropy:
\beq
\dot S = G_{ij} \beta^i \beta^j 
\label{fourb}
\eeq
which is positive semi-definite. If any of the renormalization functions $\beta^i
\not= 0$, entropy will necessarily grow. Whether the entropy might have grown
during a primordial epoch sufficiently rapidly to have generated the observed
entropy in the Universe is a dynamical question.

The most appropriate equation for addressing this dynamical question may be the
Fokker-Planck equation corresponding to (\ref{one}) 
and (\ref{four})~\cite{chaotic,emninfl}:
\beq
\partial_t {\cal P}(g, t) =\frac{1}{8\pi^2} \frac{\delta}{\delta g^i}
Q^3 \delta^{ij}\frac{\delta}{\delta g^j}\left[Q^3 {\cal P}(g,t)\right]+
\frac{\delta}{\delta g^i}\left[\beta^i {\cal P}(g, t)\right]
\label{six}
\eeq
modulo quantum ordering ambiguities for the $g$-dependent
diffusion coeeficients $Q$. The quantity 
${\cal P}(g,t)$ denotes the probability distribution in the 
theory space of strings, i.e., the probability of finding the system
in a configuration $\{ g^i \}$ at time $t$.  
The quantity 
$Q^2$ denotes the central charge
deficit of the non-critical Liouville theory, which is a measure of 
the deviation of the theory from criticality.
In our context such deviations arise by, e.g., splitting the field modes
into those inside the particle horizon in a cosmological context, 
and those that lie outside and hence are essentially unobservable
by low-energy local observers.
A stochastic equation like (\ref{six}) arises formally when one
splits a field mode $g^i$ at a time instant $t+\delta t$ as follows: 
\begin{equation}\label{sixb} 
g^i(t + \delta t) = g_c^i(t) + {\dot g}_c^i(t)\delta t + \delta_q g^i(\delta t)
\end{equation}  
The first two terms 
in (\ref{sixb}) describe the conventional classical motion of an
inflaton field  determined by the gradient of an effective potential, and the
remaining term in (\ref{sixb}) corresponds 
to quantum fluctuations in its motion,
which play a key role in chaotic inflationary models~\cite{chaotic}, 
for example. 
The average of the quantum fluctuations is connected to the diffusion 
coefficient $Q^6$ as follows: 
$< (\delta_q g)^2 > =\frac{Q^6}{4\pi^2}\delta t$.

The inflationary paradigm is now being subjected to observational tests of
ever-increasing precision~\cite{cmb}. Not only has the first acoustic peak
been measured accurately, but the second peak has been discovered and
there is good evidence for a third~\cite{cmb}. In addition to the mass
density and other cosmological parameters, the spectral index of the
inflationary perturbations is highly constrained and increasingly
sophisticated tests of Gaussianity are being applied. The conventional
field theories derived from string theory provide no convincing cancidate
for an inflaton, and the inflationary scale may not be many decades away
from the scale of quantum gravity. There is a need for a \$tring rethink
of the inflationary paradigm, and measurements of the cosmic microwave
background radiation might open an observational window on string theory. 

Another window is perhaps being provided by the increasing evidence for
cosmological vacuum energy~\cite{cmb}.
As an explicit example how this may arise in non-critical \$tring theory,
we consider the model proposed in~\cite{cosmoconst}, in which our
four-dimensional world
is viewed as a three-brane `punctured' with heavy, quantum-fluctuating
$D$ particles. As discussed in~\cite{cosmoconst}, in the context of
\$tring theory, these induce a
background physical metric
of the following Friedmann-Robertson-Walker form:  
\begin{equation}\label{physmetr}
G_{00} = -a^2, \qquad  
G_{ij} =t^4 \delta_{ij} 
\end{equation}
There is also a non-trivial dilaton field, which 
at times larger than any spatial 
scale in the problem assumes the form~\cite{cosmoconst}:
\begin{equation}\label{dilaton} 
\Phi (t) = -2{\rm ln} \, t
\end{equation} 
In general, for subasymptotic times,
the dilaton also exhibits a complicated spatial dependence, but 
for cosmological models the time dependence (\ref{dilaton}) is sufficient.

It can be readily shown that (\ref{physmetr}),(\ref{dilaton})
are solutions to Einstein and dilaton equations of motion obtained
from the following effective action (in Einstein frame) with a non-trivial 
vacuum-energy term 
\begin{equation} 
{\cal S}=\int d^4 x \sqrt{G} \left(-R + (\nabla _\mu \Phi )^2 + 
\xi e^{\zeta \Phi}\right)
\label{action4d}
\end{equation}
where $\zeta =1$, and 
\begin{equation}\label{cosmoc} 
\xi e^{\zeta \Phi} = \frac{20}{a^2 t^2}M_s^4~, 
\qquad \xi=\frac{20}{a^2}M_s^4.
\end{equation} 
Here $M_s$ is the string scale and, in our sign conventions,
the positive sign of the vacuum energy corresponds to a de Sitter-type
Universe. The normalization constant $a^2$ cannot be determined 
in this framework, and is considered as a free parameter~\footnote{The
constancy of $\xi$ is consistent with
Lorentz invariance: all the time dependence of the vacuum energy 
being due to the dilaton field $\Phi (t)$.}.

There is an alternative, more stringy way, however, to interpret the 
results (\ref{physmetr}),(\ref{dilaton}), which is appropriate
for a non-critical string framework.
One may consider the metric and dilaton functions (\ref{physmetr}),
(\ref{dilaton}) as couplings in a non-critical stringy $\sigma$-model 
with central charge deficit $Q^2$. 
As we shall see below, this allows 
the normalization constant $a^2$ to be determined 
by consistency with the generalized conformal invariance 
conditions of Liouville strings~\cite{ddk,tseytl,emninfl}: 
\begin{equation} \label{gcic}
{\ddot g}_i + Q{\dot g}_i = G^{00}\beta_i 
\end{equation}
where $Q$ is the square root (with either sign) 
of the central charge deficit. In the (toy) case of bosonic string,
$Q^2=\frac{1}{3}(c_{\rm total}[g] -25)$, where $c_{\rm total}[g]$ is the 
total central charge of the non-conformal `matter' theory, the
$g_i$ are the non-conformal $\sigma$-model couplings (omitting the
dilaton), 
and the $\beta^i$ are the appropriate renormalization-group
$\beta$ functions.
The renormalization-group $\beta$ functions in 
string theory are replaced by the so-called Weyl anomaly coefficients,
${\tilde \beta}_i$, 
taking into account the general coordinate diffeomorphism invariance 
of the target space of the $\sigma$ model, which is assumed
to be respected by the non-critical \$tring. 
The dilaton Weyl coefficient ${\tilde \beta}^\Phi$ is
given by the difference of the 
overall central charge deficit of the matter theory plus the 
Liouville contributions and the ghost 
contributions. In a Liouville framework, as a result of the restoration 
of conformal invariance, this difference vanishes. 
Hence for the dilaton one has the condition:
\begin{equation}\label{weyldil}
{\tilde \beta}^\Phi = c_{\rm total} + c_{\rm Liouville} - 26 = 0
\end{equation}
On the other hand, the dilaton $\beta$ function is determined in terms of
the rest of 
the $\beta$ functions by certain consistency relations
stemming from world-sheet renormalizability. In our interpretation
of the Liouville field as a local world-sheet scale~\cite{EMNreview}, 
these consistency
relations are assumed to be valid for the Liouville-dressed couplings
as well. 

The Minkowski sign of $G^{00} < 0$ on the right-hand side of (\ref{gcic})
is that found for the relevant supercritical-string case~\cite{aben}. The
overdot denotes differentiation with respect to the target time, which is
identified in our approach with the Liouville mode. This is an important
difference of our approach~\cite{emninfl} from apparently similar standard
Liouville string approaches~\cite{ddk,tseytl}, where the Liouville mode is
simply viewed as an extra dimension of the Liouville-dressed string. For
us, the non-conformal deformations are dressed by the Liouville mode, but
the overall target-space dimension remains the same under the
identification of the Liouville mode with the target time. For
consistency, this can only happen for supercritical strings, where the
Liouville mode is timelike~\cite{aben}. Previously, we have verified that
the $t \leftrightarrow \phi$ identification reproduces correctly the
metric of the two-dimensional string black hole~\cite{EMNreview}. In this
case, the fact that there are solutions of the generalized conditions
(\ref{gcic}, \ref{weyldil}) is a non-trivial consistency check on our
approach. 

In the problem at hand, $g_i \equiv \{G_{\mu\nu}$, $\mu,\nu=0,\dots 3 \}$.
For the graviton and dilaton backgrounds that we are considering here, 
the Weyl coefficients ${\tilde \beta}^i $  
are given to ${\cal O}(\alpha ')$ by
\begin{equation}\label{weylf}
{\tilde \beta}_{\mu\nu}^G = R_{\mu\nu} + 2\gamma \nabla_\mu \partial_\nu \Phi,
\qquad {\tilde \beta}^\Phi = -R + 
4\gamma ^2 (\nabla_\mu \Phi)(\nabla_\nu \Phi)G^{\mu\nu} - 4\gamma \nabla^2 
\Phi + Q^2 
\end{equation}
where $\gamma$ is an appropriate normalization factor to be determined
by consistency with the form (\ref{dilaton}). 
Using the explicit expression 
for the dilaton Weyl anomaly coefficient ${\tilde \beta}^\Phi$,
one can easily check the consistency of our approach,
in which the Liouville mode $\rho$ is identified with the 
target time $t$, respecting (\ref{weyldil}). The 
condition (\ref{weyldil}) is satisfied by the ${\tilde \beta}^\Phi$ 
appropriate for a $\sigma$ model in $d+1$ dimensions, the extra dimension
being provided by the Liouville mode $\rho$,
with $Q_{\rm total}^2=0$, as dictated 
by the restoration of the conformal invariance by the Liouville 
field~\cite{ddk}. In such a $\sigma$ model,
one has a dilaton coupling of the generic form dictated by Liouville
dynamics~\cite{ddk,aben,EMNreview}: 
$\Phi(t,{\vec x}, \rho)=\Phi(t,{\vec x}) + \frac{1}{2\gamma}Q\rho $, where
$Q$ is the central-charge deficit of the matter theory, e.g., that caused
by the quantum fluctuations (`recoil') 
of the $D$ particles in the specific example~\cite{cosmoconst} considered
here. Interpreting~\cite{EMNreview} $\rho$ as a local 
world-sheet scale, one has $\frac{d}{d\rho}Q=0$ on account of 
world-sheet renormalizability.  In this case, then, one may split 
the $d+1$-dimensional tensor quantities appearing in ${\tilde \beta}^\Phi$ 
into Liouville ($\rho$) components and the rest, taking into account 
the above properties of the dilaton. It is then seen immediately that
upon such a splitting, taking into account the 
cosmological backgrounds of interest to us here and the fact that
the metric component in the $\rho\rho$ direction is a constant, 
and, finally, identifying the Liouville mode $\rho$ with the target time $t$, 
one arrives easily at the expression for the ${\tilde \beta}^\Phi$ 
given in (\ref{weylf}), where now the central-charge deficit $Q^2$ of the 
`matter' theory has appeared explicitly.  It is this last expression, then,
which should vanish according to (\ref{weyldil}).

It can be checked that there are two non-trivial solutions to the above 
system of equations (\ref{gcic}, \ref{weyldil}, \ref{weylf}):
\begin{equation} 
\gamma =3/2, \qquad a^2=0.72, \, 14.41, \qquad 
-Q =\frac{1}{4~a^2~t}\left(\frac{22}{a^2}+12a^2\right)
\simeq \frac{13.61}{t}, \, \frac{3.03}{t}.
\end{equation}
We recall that the induced central charge deficit, $Q^2$, in our approach 
is defined with the appropriate dilaton exponential factors~\cite{EMNreview}
steming from the $\sigma$-model formalism. 
It, therefore, 
plays the r\^ole of
a vacuum energy $\Lambda $ 
in our non-critical \$tring theory~\cite{aben,EMNreview}, and
thus, in the above example, it relaxes to zero asymptotically as $t^{-2}$
for large times, where the analysis is valid~\cite{cosmoconst}:
\begin{eqnarray}\label{lambdacosm} 
\Lambda &=& Q^2 \sim 185 \frac{M_s^4}{t^2}~~~~ {\rm for}~~a^2=0.72, \nonumber \\
\Lambda &=& Q^2 \sim 9 \frac{M_s^4}{t^2}~~~~ {\rm for}~~a^2=14.41. 
\end{eqnarray} 
The same scaling has been found in the Einstein frame approach 
above (\ref{cosmoc}), and this justifies this 
identification of $Q^2$ with $\Lambda$.
Moreover, although the 
equations (\ref{gcic}),(\ref{weyldil}) are formally different from 
the conventional equations encountered in 
Einsteinian gravity, the $\sigma$-model metric (\ref{physmetr}) 
we find is of Robertson-Walker form, and may be identified 
with the physical metric~\footnote{We note
that the set 
of equations (\ref{gcic}) satisfies the Helmholtz
conditions~\cite{emninfl}, and so can be
derived from an {\it off-shell} action 
subject to canonical quantization. However, they characterize an
out-of-equilibrium system that does not lead to a well-defined S-matrix.}.
For $M_s \sim \frac{1}{10}M_P \sim 10^{18}$ GeV, and $t\sim 10^{60}$ in 
Planck time units, the results (\ref{cosmoc}),(\ref{lambdacosm}) 
are consistent with the current phenomenology~\cite{cmb}.  

The metric (\ref{physmetr}) has a Friedman-Robertson-Waker form
with scale factor $a(t)^2=t^4$ and 
spatial curvature $k=0$, and, as we have shown,
satisfies Einstein's equations with a scalar (dilaton) field 
$\phi$ responsible for the appearance of a cosmological constant.   
The proper
horizon distance in such a spatially-flat universe is given by:
\begin{equation}\label{horizon}
\delta_H (t) =a(t) c\int _{t_0}^\infty \frac{dt'}{a(t')} 
=ct^2/t_0^2
\end{equation}
where $c$ is the speed of light {\it in vacuo}. This is constant in the
specific model considered here, but we recall that there are 
Liouville-string based models in which there is a time-dependent
$c(t)$~\cite{emnspeed}. We observe
from (\ref{horizon}) that the expression for 
$\delta_H (t)$ is finite, and hence there is a future horizon.
This implies that the $S$-matrix description of particle scattering
breaks down in such a Universe, which is consistent with the
generic Liouville framework, as discussed previously.

It is interesting to
estimate the possible magnitude of non-Hamiltonian terms in the evolution
equation (\ref{one}) today. We have previously suggested a maximum possible
magnitude~\cite{EMNreview,adler} 
\beq
|\nd{\delta H} | \sim {E^2\over m_P}
\label{seven}
\eeq
for such terms originating from microscopic space-time foam effects. The
experimental sensitivity to tests of quantum mechanics in 
the neutral-kaon system
approaches the possible magnitude (\ref{seven})~\cite{CPLEAR}. In the case
of effects
associated with the macroscopic event horizon, we note that an event horizon of
the scale of the observable Universe would emit De Sitter radiation with a
temperature
\beq
T \sim H \sim 10^{-35} {\rm eV} \sim 10^{-31} {\rm K}
\label{eight}
\eeq
which appears somewhat distant from experimental possibilities, since the
microwave background radiation itself has a temperature $\simeq 2.7$~K,
and the
background neutrino radiation is expected to have $T \sim 1.9$~K. Thus,
the
possibility of a macroscopic event horizon today, whilst a profound theoretical
challenge, may not be of much experimental significance\footnote{It has been 
suggested~\cite{moffat} that models with a
time-varying speed of light~\cite{speed1} may remove the cosmological
horizon in an accelerating Universes, obviating the
concerns~\cite{susskind,fischler}.  We recall that such a time-varying
effective speed of light arises quite naturally in Liouville \$tring
theory, where it was proposed as a way to avoid the horizon
problem in
inflationary models~\cite{emnspeed}, and later revived 
by~\cite{speed}.  However, in our Liouville context, a
time-varying speed of light cannot save the $S$ matrix.}.

As explained in this paper, our response to this theoretical challenge is
rooted in non-critical string theory, or \$tring theory~\cite{EMNreview}. 
We had argued previously that this was the most appropriate dynamical
framework for the effective low-energy theory obtained by integrating out
higher string modes in a renormalization-group approach, and had applied
it to cosmology~\cite{emninfl,cosmoconst}. In the \$tring framework, the
basic dynamical object is the density matrix, the $S$ matrix is abandoned
in favour of the superscattering operator \$ $\not=
SS^\dagger$~\cite{Hawking}, and canonical Hamiltonian evolution is
supplemented by a non-Hamiltonian term in the quantum Liouville equation
reminiscent of open quantum-mechanical systems~\cite{ehns}. As we have
shown in this paper, \$tring theory can certainly accommodate non-zero
vacuum energy, and might even generate a time-varying value {\`a la}
quintessence.

\section*{Acknowledgements} 

J.E. thanks Farhad Ardalan for an encouraging discussion.
The work of D.V.N. is partially supported by DOE grant
DE-F-G03-95-ER-40917.  N.E.M. and D.V.N. also thank H. Hofer for his
interest and support.

\end{document}